\documentclass[conference]{IEEEtran}

\usepackage{amsfonts}
\usepackage{amssymb}
\usepackage{stfloats}
\usepackage{cite}
\usepackage{graphicx}
\usepackage{psfrag}
\usepackage{subfigure}
\usepackage{amsmath}
\usepackage{array}
\usepackage{algorithm}
\usepackage{algpseudocode}
\usepackage{epstopdf}

\begin{document}

\title{Sub-Spectrogram Segmentation for Environmental Sound Classification via Convolutional Recurrent Neural Network and Score Level Fusion}

\author{Tianhao Qiao, Shunqing Zhang, Zhichao Zhang, Shan Cao, Shugong Xu \\
Shanghai Institute for Advanced Communication and Data Science \\
Shanghai University, Shanghai, 200444, China\\
Email:\{qiaotianhao, shunqing, zhichaozhang, cshan, shugong\}@shu.edu.cn}

\maketitle

\begin{abstract}
Environmental Sound Classification (ESC) is an important and challenging problem, and feature representation is a critical and even decisive factor in ESC. Feature representation ability directly affects the accuracy of sound classification. Therefore, the ESC performance is heavily dependent on the effectiveness of representative features extracted from the environmental sounds. In this paper, we propose a sub-spectrogram segmentation based ESC classification framework. In addition, we adopt the proposed Convolutional Recurrent Neural Network (CRNN) and score level fusion to jointly improve the classification accuracy. Extensive truncation schemes are evaluated to find the optimal number and the corresponding band ranges of sub-spectrograms. Based on the numerical experiments, the proposed framework can achieve 81.9\% ESC classification accuracy on the public dataset ESC-50, which provides 9.1\% accuracy improvement over traditional baseline schemes.
\end{abstract}

\section{Introduction} \label{sect:intro}
Environmental sound classification (ESC), which automatically analyze and recognize environmental audio signals, has been widely used in surveillance, home automation, scene analysis and machine hearing\cite{barchiesi2015acoustic}.  Different from traditional sound classification, such as music and speech recognition\cite{zatorre2002structure}, ESC needs to deal with a wide range of frequency spectrum, various sounding sources, non-stationary characteristic, and noise-like signals, which triggers numerous research efforts\cite{chachada2014environmental,chu2009environmental,cowling2003comparison} recently.

Since traditional sound classification methods usually compose of {\em feature extraction} and {\em feature-based classification} procedures, a natural extension to design ESC schemes is to build more powerful components on top of the previous framework. In the feature extraction part, for example, zero crossing rate, audio tone, short-time energy \cite{jalil2013short} are commonly adopted to some sound classes in the low noise environments. However, this type of time domain feature representations often require significant computational complexity (e.g. sufficient long window size) to maintain a reasonable classification accuracy. To address this issue and fully exploit the recent progresses in image processing tasks, extracting features in the frequency domain and representing the environmental sound using time-frequency spectrogram image have been widely used\cite{dennis2014sound}. Typical frequency domain features include Mel-frequency cepstrum coefficient (MFCC)\cite{rabiner1993fundamentals} and log Mel spectrogram (Logmel)\cite{piczak2015environmental}. In the feature-based classification part, typical algorithms have undergone a paradigm shift from supervised learning (e.g., K-nearest neighbors \cite{keller1985fuzzy} and support vector machine \cite{scholkopf2001learning}) to unsupervised learning (e.g., random forest \cite{breiman2001random} and Gaussian mixture model \cite{atrey2006audio}) during the first decade of this century. In recent years, with the development of supervised learning technologies, more powerful tools have been applied to develop the feature-based classification algorithms, such as dictionary learning \cite{chu2009environmental}, matrix factorization \cite{bisot2017feature}, and deep neural networks (DNN) \cite{tokozume2018learning}.

By combining different feature extraction and feature-based classification methods, the achievable classification accuracy has been refreshed many times during recent years \cite{mcloughlin2015robust,boddapati2017classifying,piczak2015environmental,zhang2017dilated}. For example, MFCC with Multi-Layer Perception (MLP) and convolutional neural network (CNN), two different types of DNN, reached 44.9\% and 53.1\% in the public dataset ESC-50, respectively, while Logmel with CNN, is able to achieve 73.2\% in the same dataset. However, the above results simply adopt brute force methods to search the best combination, while the specific domain knowledge has not been utilized to improve the classification accuracy. Some of them are elaborated as below.

\begin{itemize}
    \item {\em Sub-spectrogram Segmentation:}
    The spectrum of frequencies of environmental sound is concentrated in low-frequency portion and a more careful study of spectrogram is usually required. As shown in \cite{phaye2018subspectralnet}, sub-spectrogram segmentation is an efficient way to improve the acoustic scene classification accuracy and whether this type of mechanism can be applied to ESC tasks is still an open question. Meanwhile, how many sub-spectrogram segments are needed for ESC tasks and how to truncate the corresponding spectrogram in the noise-like ESC environment are still challenging based on the existing literature.
    \item {\em Recurrent Architecture with Data Augmentation:} The sound is generally correlated in the time domain, and the previous frame also has an impact on the prediction of the next frame. For example, $dog\ bark$ is a sound that continues many frames. Therefore, another possible approach to improve the classification accuracy is to explore the correlations among different scales of sequence as proposed in \cite{chung2014empirical}. However, as this type of method usually requires large amount of data, an effective way to expand the limited ESC dataset, such as mixup \cite{zhang2018deep}, needs to be jointly considered.
\end{itemize}

In this paper, we propose a sub-spectrogram segmentation based ESC classification framework by jointly considering the above domain knowledge. To be more specific, we truncate the whole spectrogram into different sub-spectrograms, explore different recurrent architectures to analyze different sound features, and adopt a score level based fusion mechanism to jointly improve the classification accuracy. Extensive truncation schemes are evaluated to find the optimal number and the corresponding band ranges of sub-spectrograms. Based on the numerical experiments, the proposed framework can achieve 81.9\% ESC classification accuracy on a public dataset ESC-50, which provides 9.1\% accuracy improvement over traditional baseline schemes.

The rest of the paper is organized as follows. In Section~\ref{sect:pre}, we provide some preliminary information of Logmel spectrogram and different types of DNN. The proposed sub-spectrum segmentation based ESC classification framework is introduced in Section~\ref{sect:method} and the numerical experiments are demonstrated in Section~\ref{sect:expe}. Finally, Section~\ref{sect:conc} concludes this paper.

\section{Preliminary} \label{sect:pre}
In this section, we introduce the famous Logmel spectrogram and briefly elaborate different types of DNNs.

\subsection{Logmel Spectrogram}

\begin{figure}[h]
\centering
\includegraphics[width =3.4 in]{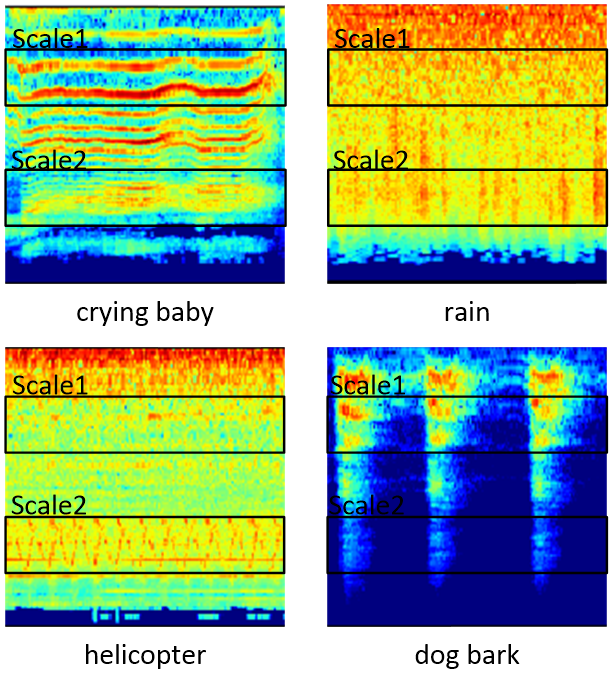}
\caption{Logmel spectrograms of \emph{crying baby, rain, helicopter and dog bark} for entire freaquency band, where the horizontal axis denotes the time dimension and the vertical axis denotes the frequency dimension.} \label{fig:feat}
\end{figure}

Consider a sampled time domain audio signal $s(t)$ and the energy spectrum density after $T$ point discrete short-time Fourier transform (STFT), $\left|S(m,n)\right|^2$, is given by,
\begin{eqnarray}\label{eqn:logmel1}
\left|S(m,n)\right|^2 = \left|\sum_{t=\frac{nT}{2}+1}^{\frac{(n+2)T}{2}}{s(t)\cdot{e^{-\frac{j2\pi{mt}}{T}}}}\right|^2,
\end{eqnarray}
for all $m\in \left[1,T/2\right]$, $n\in\left[0,N-1\right]$. The Logmel spectrogram is simply determined by applying $K$ mel-filter banks to the energy spectrum, and the corresponding mathematical expression is,
\begin{eqnarray}\label{eqn:logmel2}
S_{LM}(n,k) =
\log \left( \sum_{m=1}^{T/2}{\left|S(m,n)\right|^2}\cdot{H(m,k)} \right),
\end{eqnarray}
for $k \in \left[1, K\right]$, where $H(m,k)$ denotes the frequency response of the $k^{th}$ mel-filter in the $m^{th}$ sub-band. In the practical systems, $\{H(m,k)\}$ are selected according to the entire frequency band, $(f_L, f_H)$, e.g. from zero to half of the sampling frequency as shown in \cite{Muda2010}. In practice, the Logmel spectrogram often contains its delta and delta-delta information to form a three channel tensor, e.g., ${\mathbf{S}_{LM}(n,k)}$, according to \cite{piczak2015esc}, and some typical Logmel spectrograms of four sound categories are depicted in Fig.~\ref{fig:feat}.

\subsection{DNN}

Generally speaking, DNN is realized by connecting multiple layers of neurons to form a more powerful neural network, which typically includes MLP, CNN and recurrent neural network (RNN) structures. MLP is a classical feedforward neural network, where neurons within each layer are isolated and neurons across neighboring layers are fully connected. In the CNN architecture, the design philosophy is more or less the same, except that neurons across neighboring layers are connected via convolutional kernels and pooling operations. With the help of convolutional architecture, CNN is capable of learning local patterns among different input elements, such as image pixels or environmental sound spectrogram. The above two structures do not consider the temporal correlation among different input vectors or patterns. To address this issue, RNN is proposed to make use of previous frame-level features and learn complex temporal dynamics. By combining different architectures together, DNN has been utilized to deal with challenging tasks in ESC and other computer vision areas.

\section{Proposed Sub-spectrogram Segmentation based Classification Framework} \label{sect:method}

In this section, we introduce the proposed sub-spectrogram segmentation based classification framework, which consists of sub-spectrogram segmentation based feature extraction, CRNN based classification, and score level fusion.
\begin{figure*}[ht]
\centering
\includegraphics[width = 7.1 in]{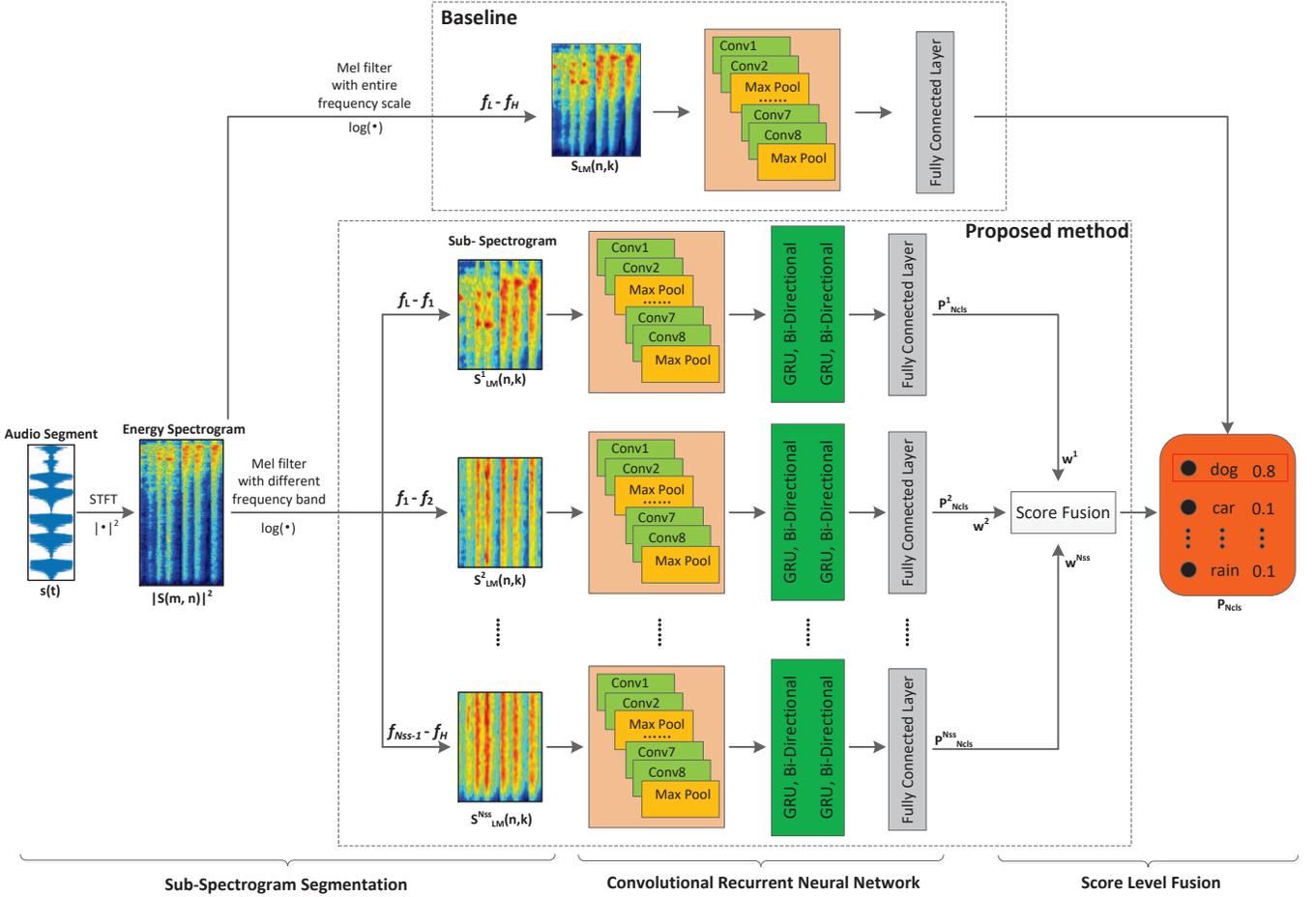}
\caption{Compare the baseline system and the proposed Sub-Spectrogram Segmentation system. In this figure, the first branch is the baseline system, which extracts Logmel features on the entire frequency band, and the other branch denotes the proposed sub-spectrogram segmentation method, which extracts Logmel on several sub-frequency bands as illustrated.}\label{fig:flow}
\end{figure*}

\subsection{Overview}

Generally speaking, the ESC task is to identify different sound categories based on the observed sound signal $s(t)$ or the equivalent energy spectrum $\left|S(m,n)\right|^2$. Given the total number of sound categories, $N_{cls}$, the mathematical expression of the classification task is,
\begin{eqnarray}
\mathbf{p}_{N_{cls}} = F\left(\left\{\left|S(m,n)\right|^2\right\} \right),
\end{eqnarray}
where $\mathbf{p}_{N_{cls}} = [p_1, p_2, \ldots, p_{N_{cls}}]^{T}$ denotes the probability distribution across $N_{cls}$ sound categories. In the traditional approach, the nonlinear function $F(\cdot)$ is directly approximated via its equivalent Logmel spectrogram and the corresponding neural network defined by $\theta$, e.g., $\mathbf{p}_{N_{cls}} = G \left(\left\{\mathbf{S}_{LM}(n,k) \right\};\theta \right)$.

An overview of the proposed sub-spectrgram segmentation based classification framework is shown in Fig.~\ref{fig:flow}. Instead of generating the Logmel spectrogram based on the entire frequency band as elaborated before, we truncate the whole spectrogram into $N_{ss}$ parts, e.g. $(f_L,f_1), \ldots, (f_{N_{ss} - 1}, f_H)$, and perform the decision based on a score level fusion. Mathematically, the overall operations can be described through,
\begin{eqnarray}
\mathbf{p}_{N_{cls}} = \sum_{i = 1}^{N_{ss}} \omega^{i} \mathbf{p}_{N_{cls}}^{i} = \sum_{i = 1}^{N_{ss}} \omega^{i} G \left(\left\{\mathbf{S}^{i}_{LM}(n,k) \right\};\theta \right),
\end{eqnarray}
where $\omega^{i}$ and $\mathbf{p}_{N_{cls}}^{i}$ denote the fusion weight and the score of the $i^{th}$ sub-spectrogram, respectively, and $\sum_{i = 1}^{N_{ss}} \omega^{i} = 1$. $\left\{\mathbf{S}^{i}_{LM}(n,k) \right\}$ defines the generated Logmel spectrograms based on the $i^{th}$ band\footnote{For illustration purpose, we define $f_0 = f_L$ and $f_{N_{ss}} = f_H$.}, e.g., from $f_{i-1}$ to $f_i$, and $G(\cdot;\theta)$ represents a nonlinear mapping relation between the Logmel spectrogram and the classification result.

\subsection{Sub-Spectrogram Segmentation}\label{sub}

\begin{figure}[ht]
\centering
\includegraphics[width =3.4 in]{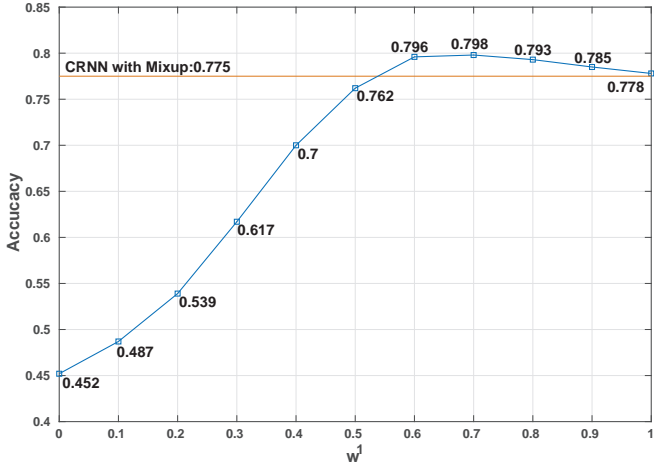}
\caption{Compare using and not using sub-spectrogram segmentation on CRNN with Mixup system. The blue line indicates the classification accuracy of different weights cases, while the orange line indicates the accuracy of CRNN with Mixup system.} \label{fig:curve}
\end{figure}

As shown in Fig.~\ref{fig:feat}, different scales of the spectrogram may behave significantly different. Inspired by this phenomenon, we simply partition the whole spectrogram into two parts, e.g., $N_{ss}=2$ with $f_L = 0$ kHz , $f_1 = 10$ kHz, and $f_H = 22.05$ kHz, to demonstrate the effectiveness of the proposed sub-spectrogram segmentation based ESC scheme. The tested classification accuracy versus $\omega^1$ relation on ESC-50 dataset\cite{piczak2015esc} is shown in Fig.~\ref{fig:curve}. From this figure, we can conclude that the proposed sub-spectrogram segmentation based classification scheme with proper weight assignment can outperform the baseline system.

The second task to design an effective solution for ESC is to identify the optimized number of sub-spectrogram segments, e.g., $N_{ss}^{\star}$. Although the rigorous mathematical derivation is still challenging, we provide extensive numerical studies to obtain the result. Specifically, we evaluate the classification accuracy under different values of $N_{ss}$ and generate the segmentation points, $\{f_i\}$, accordingly. With the optimized weight coefficients, $\{\omega^{i}\}$, the tested results are listed in Table~\ref{tab:sub}. As shown in Table~\ref{tab:sub}, the classification accuracy is {\em NOT} monotonically increasing with respect to $N_{ss}$ and the optimized number of sub-spectrogram segments in this case is $N_{ss}^{\star}=4$.

\begin{table}[ht]
\caption{Classification accuracy under different values of $N_{ss}$.
\label{tab:sub}}
\centering
\setlength{\tabcolsep}{3.5mm}{
\begin{tabular}{ccccccccc}
\hline
\hline
$N_{ss}$ &$f_L$(kHz) &$\{f_i\}$(kHz) &$f_H$(kHz) & Accuracy\\
\hline
1 &0 &- &22.05 &77.5\% \\
\hline
2 &0 &10 &22.05 &79.8\% \\
\hline
3 &0 &6,10 &22.05 &81.6\% \\
\hline
4 &0 &3,6,10 &22.05 &\textbf{81.9\%} \\
\hline
5 &0 &3,6,10,15 &22.05 &81.3\% \\
\hline
6 &0 &3,6,10,13,16 &22.05 &81.1\% \\
\hline
\hline
\end{tabular}}
\end{table}

\subsection{CRNN with Mixup}\label{CRNN with Mixup}

In the field of ESC, CNN is capable of capturing local spectro-temporal patterns by using convolution kernels with a small receptive field on spectrogram features, and RNN has the ability of learning temporal relationships for environmental sound features. Motivated by the fact that CNN and RNN are complimentary in their modeling capability, we combine them in a unified architecture, named {\em convolutional recurrent neural network (CRNN)}, to approximate the original nonlinear function $G(\cdot;\theta)$, where the detailed architecture and parameters of CRNN are presented in Table~\ref{tab:crnn}. Specifically, the learned features through conventional CNN architectures are forwarded to bi-directional gated recurrent unit (GRU) for temporal processing before obtaining the final score of the $i^{th}$ sub-spectrogram, $\mathbf{p}_{N_{cls}}^{i}$.

\begin{table}[ht]
\caption{Architecture of the Proposed Convolutional Recurrent Neural Network (CRNN)
\label{tab:crnn}}
\centering
\setlength{\tabcolsep}{2mm}{
\begin{tabular}{ccccccccc}
\hline
Layer &Nums of filters &Filter Size &Stride &Output Size\\
\hline
Conv1 &32 &(3, 3) &(1,1) &(60,60,32)\\
Conv2 &32 &(3, 3) &(1,1) &(60,60,32)\\
Pool1 &- &- &(4, 2) &(15,30,32)\\
\hline
Conv3 &64 &(3, 1) &(1, 1) &(15,30,64)\\
Conv4 &64 &(3, 1) &(1, 1) &(15,30,64)\\
Pool2 &- &- &(2, 1) &(8,30,64)\\
\hline
Conv5 &128 &(1, 3) &(1, 1) &(8,30,128)\\
Conv6 &128 &(1, 3) &(1, 1) &(8,30,128)\\
Pool3 &- &- &(1, 2) &(8,15,128)\\
\hline
Conv7 &256 &(3, 3) &(1, 1) &(8,15,256)\\
Conv8 &256 &(3, 3) &(1, 1) &(8,15,256)\\
Pool4 &- &- &(2, 2) &(4,8,256)\\
\hline
GRU1 &256 &- &- &(8,256)\\
GRU2 &256 &- &- &(8,256)\\
\hline
FC1 &nums of classes &- &- &(nums of classes,  )\\
\hline
\end{tabular}}
\end{table}


To avoid possible overfitting caused by limited training data, we use Mixup data augmentation method to construct virtual training data and extend the training distribution \cite{zhang2017mixup}. In mixup, each virtual training data is generated by mixing two training samples, e.g., mixing a $dog\ bark$ Logmel spectrogram and a $crying\ baby$ Logmel spectrogram to get a mixed feature, the formula for mixing is determined by

\begin{equation}\label{eqn:mapping}
\{\widetilde{S}^i_{LM}(n,k)\}= {\lambda}\{S^i_{LM}(n,k)\}_j + (1-\lambda)\{S^i_{LM}(n,k)\}_{j'},
\end{equation}
where $\{S^i_{LM}(n,k)\}_j$ and $\{S^i_{LM}(n,k)\}_{j'}$ are Logmel spectrograms of two samples randomly selected from training data. Since the features are mixed, the labels corresponding to each sound should be mixed, and the class label used for the mixed samples are generated with the same proportion. In addition, the mix factor $\lambda$ is decided by a hyper-parameter $\alpha$ and $\lambda$ $\sim$ Beta($\alpha$, $\alpha$) \cite{zhang2017mixup}.

As shown in TABLE \ref{tab:crnn mixup}, when using CRNN or mixup, the accuracy can be improved by 2.3\% and 3\% respectively, and when using CRNN with mixup, the classification accuracy is 4.7\% more than it of the baseline system.

\begin{table}[ht]
\caption{Classification accuracy of whether to use CRNN or mixup.
\label{tab:crnn mixup}}
\centering
\setlength{\tabcolsep}{10.1mm}{
\begin{tabular}{ccccccccc}
\hline
\hline
Network &Mixup &Accuracy\\
\hline
CNN &$\times$ &72.8\% \\
\hline
CRNN &$\times$ &75.1\% \\
\hline
CNN &$\surd$ &75.8\% \\
\hline
CRNN &$\surd$ &77.5\% \\
\hline
\hline
\end{tabular}}
\end{table}

\subsection{Score Level Fusion}\label{Score Level Fusion}

After determining the optimized number of sub-spectrogram segments $N_{ss}^{\star}$ and an approximation of $G(\cdot;\theta)$ using CRNN, the remaining task is to identify the optimized weights in the score level fusion, e.g., $\{\omega^{\star,i}\}$. By exhaustively searching over all the possible combinations of $\{\omega^{i}\}$, we can obtain the optimized weights and the accuracy results as shown in Table~\ref{tab:score}. Based on the experimental results, the score level fusion provides 2.5\% to 3.7\% accuracy improvement if compared with uniform weights assignment.

\begin{table}[ht]
\caption{Classification Accuracy under Different Score Level Fusion Strategies}
\label{tab:score}
\centering
\setlength{\tabcolsep}{2.7mm}{
\begin{tabular}{ccccccccc}
\hline
\hline
$N_{ss}$ &$f_L$(kHz) &$\{f_i\}$(kHz) &$f_H$(kHz) &Fusion & Accuracy\\
\hline
2 &0 &10 &22.05 &$\times$ &76.2\% \\
\hline
2 &0 &10 &22.05 &$\surd$ &79.8\% \\
\hline
3 &0 &6,10 &22.05 &$\times$ &77.9\% \\
\hline
3 &0 &6,10 &22.05 &$\surd$ &81.6\% \\
\hline
4 &0 &3,6,10 &22.05 &$\times$ &79.4\% \\
\hline
4 &0 &3,6,10 &22.05 &$\surd$ &81.9\% \\
\hline
5 &0 &3,6,10,15 &22.05 &$\times$ &77.7\% \\
\hline
5 &0 &3,6,10,15 &22.05 &$\surd$ &81.3\% \\
\hline
\hline
\end{tabular}}
\end{table}

\section{Experiments} \label{sect:expe}

In this section, we perform some numerical experiments on a public dataset called ``ESC-50'' \cite{piczak2015esc} to demonstrate the effectiveness of the proposed sub-spectrogram segmentation scheme. In order to provide a fair comparison, all the evaluations are performed on a standard server environment with Nvidia P100 GPU.  Keras library with TensorFlow backend is installed and mini-batch stochastic gradient descent with Nesterov momentum of 0.9 is adopted to train the network models. The learning rate is initialized to $0.1$ and shrinks by $10$ times for each $100$ epochs. We choose the batch size to be 200 and the cross entropy between true labels and prediction labels is adopted as the loss function, which is commonly used for multi-classification tasks. The remaining important parameters are
listed in Table~\ref{tab:parameters}.

\begin{table}[ht]
\caption{Parameter settings in experiments.
\label{tab:parameters}}
\centering
\setlength{\tabcolsep}{6.3mm}{
\begin{tabular}{ccc}
\hline
\hline
Parameters &Definition &Values\\
\hline
$f_s$ &sampling frequency &44100\\
\hline
$N_{cls}$ &number of classes &50\\
\hline
$T$ &STFT point &1024\\
\hline
$N$ &frame length &60\\
\hline
$K$ &number of mel-filter banks &60\\
\hline
$\alpha$ &Mixup hyper-parameter &0.2\\
\hline
\hline
\end{tabular}}
\end{table}

In the following scenarios, we provide some numerical comparisons between the proposed sub-spectrogram segmentation scheme with the conventional baseline scheme, where a simple CNN architecture is used to model the relation between the entire Logmel spectrogram and the final classification results as shown in Fig.~\ref{fig:flow}. In order to obtain more accurate results, we test the classification accuracy under different $N_{ss}$, $\{f_i\}$, and $\{\omega^{i}\}$ to show the effect of sub-specrogram segmentation as well as score level fusion in what follows.

\subsection{Effect of Sub-Spectrogram Segmentation}

As shown in Table~\ref{tab:result}, we analyze these results with different factors, including $N_{ss}$, $\{f_i\}$, and $\{\omega^{i}\}$. We first choose a number of $f_i$, and then combine some of these to get a set of situations, including different $N_{ss}$ and the same $N_{ss}$ with different $\{f_i\}$. Finally we assign different $\{\omega^{i}\}$ to each case for testing the performance of the model several times.

\begin{table}[ht]
\caption{Classification accuracy under different $N_{ss}$, $\{f_i\}$, and $\{\omega^{i}\}$.
\label{tab:result}}
\centering
\setlength{\tabcolsep}{1mm}{
\begin{tabular}{ccccccccc}
\hline
\hline
$N_{ss}$ &$f_L$(kHz) &$\{f_i\}$(kHz) &$f_H$(kHz) &$\{w^i\}$ &Accuracy\\
\hline
1 &0 &- &22.05 &1 &77.5\% \\
\hline
2 &0 &10 &22.05 &0.7,0.3 &79.8\% \\
\hline
3 &0 &10,20 &22.05 &0.5,0.3,0.2 &80.0\% \\
\hline
3 &0 &7,14 &22.05 &0.5,0.2,0.3 &80.3\% \\
\hline
3 &0 &6,10 &22.05 &0.5,0.3,0.2 &81.6\% \\
\hline
4 &0 &10,15,20 &22.05 &0.5,0.2,0.2,0.1 &80.2\% \\
\hline
4 &0 &5,10,15 &22.05 &0.4,0.3,0.1,0.2 &80.7\% \\
\hline
4 &0 &3,6,10 &22.05 &0.4,0.2,0.2,0.2 &\textbf{81.9\%} \\
\hline
5 &0 &10,13,16,19 &22.05 &0.4,0.2,0.1,0.2,0.1 &80.6\% \\
\hline
5 &0 &5,10,15,20 &22.05 &0.4,0.2,0.1,0.2,0.1 &80.5\% \\
\hline
5 &0 &3,6,10,15 &22.05 &0.4,0.2,0.2,0.1,0.1 &81.3\% \\
\hline
6 &0 &3,6,10,13,16 &22.05 &0.3,0.2,0.2,0.1,0.1,0.1 &81.1\% \\
\hline
6 &0 &6,10,13,16,19 &22.05 &0.4,0.1,0.2,0.1,0.1,0.1 &81.2\% \\
\hline
\hline
\end{tabular}}
\end{table}

The effect of $N_{ss}$ has been analyzed in \ref{sub}, and the optimized number of sub-spectrogram segments is $N_{ss}^{\star}=4$. Then, we compare the classification accuracy with different $\{f_i\}$. We experimented three selection ways of $\{f_i\}$, including more segments for low sub-frequency portion, approximately average segmentation and more segments for high sub-frequency portion, the results are recorded in TABLE \ref{tab:result} from top to bottom, respectively. The result shows that we obtain the higher classification accuracy than others when applying more segments for low sub-frequency portion.

By analyzing the curve changing in Fig.\ref{fig:curve}, we can see that the low sub-frequency band contains most of the characteristics of the environmental sounds, and although the high sub-frequency band contains few of the characteristics, it is still indispensable for the classification. According to this, we try to appropriately increase the $\{\omega^{i}\}$ of the low sub-frequency segment during the fusion and find that we can get better results. The $\{\omega^{i}\}$ listed in TABLE \ref{tab:result} are optimal in the corresponding case, and the $\{\omega^{i}\}$ of low sub-frequency portion is generally higher than the $\{\omega^{i}\}$ of high sub-frequency portion.

\subsection{Effect of the proposed method}

We further compared the combination of CRNN, Mixup, segmentation and fusion. As shown in TABLE \ref{tab:crnn mixup sub}, the classification accuracy is improved when using CRNN, Mixup, segmentation and score level fusion. Specifically, we obtained the highest accuracy of 81.9\% with the combination of these methods and boosted an absolutely improvement of 9.1\% than baseline system, which demonstrated the effectiveness of proposed methods.

\begin{table}[ht]
\caption{Comparison for combination of CNN, RNN, Miuxp, segmentation and fusion. When using segmentation, the $N_{ss}$, $\{f_i\}$ and $\{\omega^{i}\}$ are set as 4, \{3, 6, 10\} and \{0.4, 0.2, 0.2, 0.2\}, respectively.
\label{tab:crnn mixup sub}}
\centering
\setlength{\tabcolsep}{3.7mm}{
\begin{tabular}{ccccccccc}
\hline
\hline
Network &Mixup &Segmentation &Fusion &Accuracy\\
\hline
CNN &$\times$ &$\times$ &$\times$ &72.8\% \\
\hline
CRNN &$\times$ &$\times$ &$\times$ &75.1\% \\
\hline
CNN &$\surd$ &$\times$ &$\times$ &75.8\% \\
\hline
CRNN &$\surd$ &$\times$ &$\times$ &77.5\% \\
\hline
CNN &$\times$ &$\surd$ &$\times$ &76.1\% \\
\hline
CNN &$\times$ &$\surd$ &$\surd$ &77.4\% \\
\hline
CRNN &$\times$ &$\surd$ &$\times$ &78.0\% \\
\hline
CRNN &$\times$ &$\surd$ &$\surd$ &79.3\% \\
\hline
CNN &$\surd$ &$\surd$ &$\times$ &78.2\% \\
\hline
CNN &$\surd$ &$\surd$ &$\surd$ &80.7\% \\
\hline
CRNN &$\surd$ &$\surd$ &$\times$ &79.4\% \\
\hline
CRNN &$\surd$ &$\surd$ &$\surd$ &\textbf{81.9\%} \\
\hline
\hline
\end{tabular}}
\end{table}

\section{Conclusion} \label{sect:conc}
In this paper, we propose a classification framework, including the sub-spectrogram segmentation, CRNN, Mixup and score level fusion, which can improve the classification accuracy on a public dataset ESC-50. Through evaluating the extensive truncation schemes experiments, we find the optimal number of sub-spectrogram segments is 4, and either appropriately applying more segments for low sub-frequency portion or assigning higher weight to low sub-frequency portion can improve the classification accuracy. In addition, the proposed framework can achieve 81.9\% ESC classification accuracy, which provides 9.1\% accuracy improvement over traditional baseline schemes. In the future work, we will try to use neural networks to learn the selection strategy of segmentation points automatically.

\bibliographystyle{IEEEtran}
\bibliography{IEEEabrv,bb_rf}

\end{document}